\documentclass[a4paper]{jpconf}
\usepackage{amssymb}
\usepackage{amsmath}
\usepackage{iopams}
\usepackage{graphicx}
\usepackage{epsfig}

\newcommand\rb{{\bf r}}
\newcommand\xb{{\bf x}}

\begin{document}
\title{The frictional Schr\"odinger---Newton equation in models of wave function collapse}

\author{Lajos Di\'osi}

\address{Research Institute for Particle and Nuclear Physics, H-1525 Budapest 114, P.O.Box 49, Hungary}

\ead{diosi@rmki.kfki.hu}

\begin{abstract}
Replacing the Newtonian coupling $G$ by $-iG$, the Schr\"odinger---Newton equation
becomes ``frictional''. Instead of the reversible Schr\"odinger---Newton equation, 
we advocate its frictional version to generate the set of pointer states for 
macroscopic quantum bodies.
\end{abstract}

\section{Introduction}
\label{Intro}
\begin{figure}[h]
\begin{center}
\includegraphics[scale=.70]{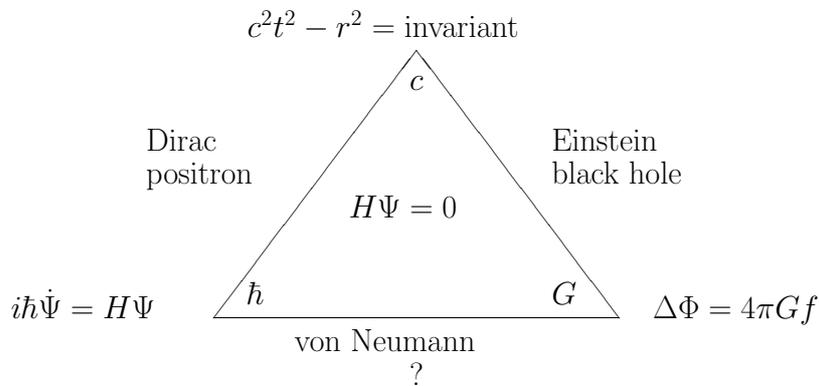}
\caption{\label{fig1}$c=$velocity of light, $G=$Newton's
gravitational constant, $\hbar=$Planck constant. The corners of the triangle
represent the three fundamental theories, the sides correspond to partially
unified theories while the middle symbolizes the fully unified theory.}
\end{center}
\end{figure}

Surely there must be a sort of bottleneck that impedes the unification
of both the quantum and the gravitation theories. In the
mainstream opinion, the bottleneck is the concept of gravity  rather
than the concept of quantum, possibly we have to reformulate our concept
of spacetime and introduce sophisticated representations of the
geometry. In a sidestream opinion (cf., e.g., \cite{Pen86,GelHar90,Dio92,Dio07,Adl04}), 
the bottleneck is within the concept of quantum theory, maybe in the notorious
concept of wave function collapse which blocks our progress to a unified quantum gravity. 
If we support this sidestream opinion, it deserves a look at 
Fig.~\ref{fig1} of the Physics Building \cite{Dio92,Dio07}. 
It might well be that
the path upto a relativistic theory of a quantized 
Universe goes through the non-relativistic theory of Newtonian Quantum 
Gravity explaining 
the quantized motion of common macroscopic objects.
If the conflict of wave function collapse with the concept of gravity
is indeed the bottleneck then, in the Newtonian limit, the resolution
should come in the form of a gravity-related (but non-Hamiltonian) 
theory of spontaneous (i.e. non-environmental) {\it decoherence and collapse} of 
macro-objects' wave function. This task is complex and maybe too elusive.
It is, therefore, in tradition that we focus on the quantized motion
of a rigid massive ball prepared in a so-called Schr\"odinger cat state
and we suppose a certain Newton-gravity-related decoherence for it.
We are going to adopt this focused discussion of the quantized motion 
of massive objects, which might eventually clear a path
to the relativistic unification of the quantum and gravity.

\section{The `rigid ball' Schr\"odinger cat and the Newtonian $U(\xb-\xb')$}
\label{sec_rigid}
Our macroscopic system is a rigid homogenous ball.
Initially we start from a Schr\"odinger cat state. It means that the c.o.m. wave function 
of the ball is a balanced superposition of two distant wave packets. We postulate  
a quick universal (non-environmental) decoherence-collapse mechanism which destroys 
the superposition in a short time $\sim t_G$ (decoherence), also drops one of the wave packets and retains 
the other at random (collapse). This way the initial macroscopic (i.e.: large-distance) superposition
\begin{figure}[h]
\begin{center}
\includegraphics[scale=.70]{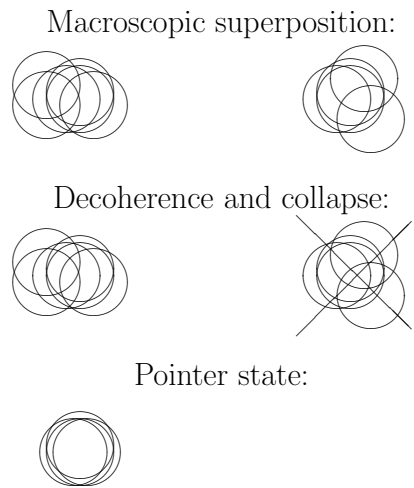}
\caption{\label{fig2}The Schr\"odinger cat state (upper part) decoheres and collapses to one of the
two wave packets (middle part) and a localized pointer state is formed ultimately (lower part).}
\end{center}
\end{figure}
has disappeared. The local superpositions (i.e.: internal to one wave packet) survive and, 
according to a further assumption, become uniform later when the wave packet takes a steady shape. 
We call this ultimate state {\it pointer state}. This qualitative picture of the process 
two-packets---one-packet---one-pointer-state, see Fig.~\ref{fig2}, if it were real, would relax the controversy
between quantum mechanics and macroscopic physics. Anyway, it deserves the concrete study:
are there simple equations at all, to model the postulated scenario? 

Due to the extreme simplification of the system, we can use transparent
equations. All equations will contain the same Newtonian structure. Formally,
this structure is just the Newtonian self-interaction potential of two 
hypothetical interpenetrating copies of our rigid ball centered at $\xb$ and
$\xb'$, respectively:
\begin{equation}
U(\xb-\xb')=
-G\int\frac{f(\rb|\xb)f(\rb'|\xb')}{\vert\rb'-\rb\vert}d\rb d\rb'~,
\end{equation}
where $G$ is the Newton constant, $f(\rb|\xb)=(3M/4\pi R^3)\theta(R-|\rb-\xb|)$ is the mass density 
at $\rb$ of the homogeneous rigid ball of mass $M$ and radius $R$, $\theta$ stands for the step function.
This potential takes the standard elementary form for large separations and it becomes
a quadratic potential for small separations \cite{Dio84}:  
\begin{equation}
U(\xb-\xb')\sim
\left\{\begin{array}{cl}-GM^2/|\xb-\xb'|                           &\mathrm{for~}|\xb-\xb'|\gg R\\
                                                                 &\\
                        U(0)+\frac{1}{2}M\omega_G^2|\xb-\xb'|^2~~~~&\mathrm{for~}|\xb-\xb'|\ll R
       \end{array}\right.
\end{equation}
where $\omega_G^2=GM/R^3$ is the frequency of the corresponding hypothetic oscillator.
This $U(\xb-\xb')$ will be utilized in various equations below, namely, in the equation of
the c.o.m. decoherence, in the equation of the pointer states, and in the equation of both.

\section{The equation of c.o.m. decoherence time}
\label{sec_decoh}
Before we suggest any differential equation for the process of decoherence, we restrict ourselves
to the equation of the decoherence time $t_G$ which is needed to collapse the macroscopic superposition,
cf. the step from the upper stage to the middle one on Fig.~2. 
The suggested general expression reads \cite{Dio87a,Dio89,Pen94,Pen96}:
\begin{equation}\label{t_G}
t_G=:\frac{\hbar}{U(\xb-\xb')-U(0)}~.
\end{equation}
For distant superposition, like the Schr\"odinger cat state on Fig.~2, we get:
\begin{equation}
t_G\sim-\hbar/U(0) \sim \hbar R/GM^2~.
\end{equation}
For atomic masses, $t_G$ is extremely long and the corresponding decoherence effect is irrelevant.
For nano-objects, $t_G$ is shorter and the postulated universal decoherence may compete with the
inevitable environmental decoherence. This would be the regime of experimental interest. 
For large macro-objects, $t_G$ is unrealistically short, i.e., they would never form a Schr\"odinger cat
state --- which is just the consequence we desire.

Hence, the Eq.~(\ref{t_G}) yields a plausible scale of spontaneous decoherence. There is a serious
problem, however. For a point-like massive ball $(R=0)$ as well as for any object 
containing point-like massive constituents $U(0)$ is $-\infty$ therefore $t_G$ would be zero! 
This can not be physical, we can not suppose infinite fast decoherence in any case. Therefore,
the Eq.~(\ref{t_G}) can not be universally valid unless, e.g., we introduce a cutoff to resolve 
the issue that the decoherence speed $1/t_G$ diverges for point-like objects. 

Let us anticipate that the simple form (\ref{t_G}) of the decoherence-collapse time induces a
natural form for the differential equation of the detailed stochastic process. This equation
will be discussed in Sec.~\ref{sec_match}.

\section{Pointer states: SNE?}
\label{sec_SNE}
For long enough time we expect that the c.o.m. quantum state of our ball becomes a so-called
pointer state which should be localized wave packet of constant shape. Let us consider the
so-called Schr\"odinger---Newton equation (SNE) \cite{Dio84,Pen98} for the c.o.m. wave function $\psi(\xb)$:  
\begin{equation}\label{SNE}
\frac{d\psi(\xb)}{dt}=
\hbox{standard q.m. terms}-\frac{i}{\hbar}\int U(\xb-\xb')\vert\psi(\xb')\vert^2d\xb'~\psi(\xb)~.
\end{equation}
The ground state solution is a standing soliton of width $\Delta\xb_G$. Galilean translations
and boosts yield an overcomplete set of states and we identify them as the set of pointer states. 

For atomic particles, 
$\Delta\xb_G$  is extremely large, the pointer states are practically delocalized. For nano-objects,
the localization becomes significant. For a rigid ball of common density the approximation 
$\Delta\xb_G\ll R$ is valid if $R\gg 10^{-5}$cm and $M\gg 10^{-15}$g. Then 
$U(\xb-\xb')\approx U(0)+\frac{1}{2}M\omega_G^2|\xb-\xb'|^2$ and the SNE reduces to:
\begin{equation}
\frac{d\psi(\xb)}{dt}=
\hbox{standard q.m. terms}-\frac{i}{2\hbar}M\omega_G^2|\xb-\langle\xb\rangle|^2\psi(\xb)~.
\end{equation}
If there is no external potential, the exact ground state solution is easy because in the
special framework $\langle\xb\rangle=0$ the equation describes a harmonic oscillator of 
mass $M$ and frequency $\omega_G$. We can just read out the ground state wave function
of the oscillator from a textbook:
\begin{equation}\label{SNEsol}
\psi(\xb)=\mathcal{N}\exp\left(-\frac{\xb^2}{4\Delta x_G^2}\right)~,~~~~~
\Delta\xb_G=\sqrt{\frac{\hbar}{M\omega_G}}=\left(\frac{\hbar^2}{GM^3}\right)^{1/4}R^{3/4}~.
\end{equation}
 
The SNE (\ref{SNE}) is a non-linear and reversible. Contrary to the 
decoherence-collapse speed $1/t_G$, the SNE remains regular for $R=0$,
i.e., for point-like objects or constituents as well. 
The SNE is utilized to generate the set of pointer states while it is not clear
how the wave function would reduce to these pointer states and whether the SNE has any
role in the {\it process} of reduction. 

\section{Pointer states: frictional SNE!}
\label{sec_frSNE}
The SNE is not the only plausible choice to generate the pointer states. 
We are going to consider an  alternative, the frictional Schr\"odinger---Newton
equation (frSNE):  
\begin{equation}
\frac{d\psi(\xb)}{dt}=
\hbox{standard q.m. terms}-\frac{1}{\hbar}\int U(\xb-\xb')\vert\psi(\xb')\vert^2d\xb'~\psi(\xb)
                          +\frac{1}{\hbar}U_G\psi(\xb)~,
\end{equation}
where $U_G=\int\!\!\int U(\xb''-\xb')\vert\psi(\xb')\psi(\xb'')\vert^2d\xb'd\xb''$. This equation
differs from the SNE by the imaginary coupling $-iG$ of the non-linear term instead of the real
coupling $G$. A further technical difference consists of the trivial new term with $U_G$, that assures
the normalization of the state.

We expect that the ground state solution is a standing soliton of width $\Delta\xb_G$ of the same order
like for the SNE, and Galilean translations-boosts yield the overcomplete 
set of pointer states. 

For atomic particles, similarly to the SNE, 
$\Delta\xb_G$  is extremely large and can not represent any significant localization. For nano-objects
the localization becomes significant. For a rigid ball of common density the approximation 
$\Delta\xb_G\ll R$ applies if $R\gg 10^{-5}$cm and $M\gg 10^{-15}$g. Then 
$U(\xb-\xb')\approx U(0)+\frac{1}{2}M\omega_G^2|\xb-\xb'|^2$ and the frSNE reduces to:
\begin{equation}\label{frSNEquad}
\frac{d\psi(\xb)}{dt}=
\hbox{standard q.m. terms}-\frac{1}{2\hbar}M\omega_G^2|\xb-\langle\xb\rangle|^2\psi(\xb)
                          +\frac{1}{2\hbar}M\omega_G^2 \langle(\Delta\xb)^2\rangle\psi(\xb)~,
\end{equation}
where $\langle(\Delta\xb)^2\rangle=\langle\xb^2\rangle-\langle\xb\rangle^2$. 
If there is no external potential, the exact ground state solution is simple:
\begin{equation}
\psi(\xb)=\mathcal{N}\exp\left(-\sqrt{-i}\frac{\xb^2}{4\Delta x_G^2}\right),~~~~~\sqrt{-i}=\frac{1-i}{\sqrt{2}}~.
\end{equation}
As we see, the ground state is identical with the ground state (\ref{SNEsol}) of the SNE provided we
replace $G$ by $-iG$. 

The non-linear frSNE (\ref{SNE}) is, contrary to the SNE, not reversible. The frSNE keeps to be regular for $R=0$. 
When we generated the pointer states by the SNE in Sec.~\ref{sec_SNE} it was not clear 
``how the wave function would reduce to these pointer states and whether the SNE has any
role in the {\it process} of reduction''. The frSNE is on the positive side now: its role in the process
of reduction is clear. At least there is a natural choice for the stochastic dynamics, see Sec.~\ref{sec_match}.

\section{Matching decoherence-collapse with pointer states}
\label{sec_match}
As we mentioned in Sec.~\ref{sec_decoh}, there is a straightforward mathematical model \cite{Dio89} of the
decoherence-collapse process that emerges directly from the decoherence-collapse speed $1/t_G$ 
defined by (\ref{t_G}). The central structure of the model is the standard von Neumann equation 
of motion for the c.o.m. density matrix $\rho(\xb,\xb')$, completed by a non-Hamiltonian decoherence 
term $-t_G^{-1}\rho(\xb,\xb')$ \cite{Dio87a}: 
\begin{equation}\label{vNNE}
\frac{d\rho(\xb,\xb')}{dt}=
\hbox{standard q.m. terms}-\frac{1}{\hbar}[U(\xb-\xb')-U(0)]\rho(\xb,\xb')~.
\end{equation}
Obviously, the new term tends to destroy the interference terms at a time scale $t_G$ --- which was
the effect to model. This von Neumann---Newton equation (vNNE), see originally in \cite{Dio07}, 
describes decoherence but it does not yet describe collapse and the subsequent reduction to a pointer 
state either. Nonetheless, there is a natural stochastic modification (also called unraveling) of the 
vNNE which can do it:
\begin{eqnarray}\label{stvNNE}
\frac{d\rho(\xb,\xb')}{dt}=
\hbox{standard q.m. terms}&-&\frac{1}{\hbar}[U(\xb-\xb')-U(0)]\rho(\xb,\xb')\\
                          &+&\frac{1}{\hbar}[W_t(\xb)+W_t(\xb')-2\langle W_t\rangle]\rho(\xb,\xb')~,\nonumber
\end{eqnarray}
where $\langle W_t\rangle=\int W_t(\xb)\rho(\xb,\xb)d\xb$ and $W_t$ itself is a white-noise field 
with spatial correlation: $\mathrm{M}[W_t(\xb)W_{t'}(\xb')]=-\hbar U(\xb-\xb')\delta(t-t')$. 

After long time, this stochastic differential equation drives any initial state $\rho(\xb,\xb')$ into 
a localized pure state --- the pointer state --- while the stochastic differential equation reduces to the following
stochastic wave equation:
\begin{eqnarray}\label{stWE}
\frac{d\psi(\xb)}{dt}=
\hbox{standard q.m. terms}&-&\frac{1}{\hbar}\int U(\xb-\xb')\vert\psi(\xb')\vert^2d\xb'~\psi(\xb)
                             +\frac{1}{2\hbar}[U_G+U(0)]\psi(\xb)\nonumber\\
                          &+&\frac{1}{\hbar}[W_t(\xb)-\langle W_t\rangle]\psi(\xb)~.   
\end{eqnarray}
We can recognize the frSNE plus a stochastic term! (The change $U_G\rightarrow [U_G+U(0)]/2$ of the trivial 
renormalizing term compensates the contribution of the singular stochastic term.)
We are not able to solve this equation in the general case.
We can do it in the limit where the c.o.m. is already well localized with respect to the size $R$ of the ball.
In this limit the equation reads: 
\begin{equation}\label{stWEquad}
\frac{d\psi(\xb)}{dt}=
\hbox{standard q.m. terms}-\frac{1}{2\hbar}M\omega_G^2|\xb-\langle\xb\rangle|^2\psi(\xb)
                          +w_t\sqrt{\frac{M}{\hbar}}\omega_G(\xb-\langle\xb\rangle)\psi(\xb)~,
\end{equation}
where $w_t$ is standard white-noise of correlation $M[w_tw_{t'}]=\delta(t-t')$. If there is no external
potential, the exact steady shape solutions of this equation are the pointer state solutions of the
frSNE and the effect of the stochastic term only contributes to the random walk of the steady wave packet. 

\section{Context}
My work advocates the frictional alternative of the
SNE instead of the SNE itself. The latter has been studied in a number of works including physics and mathematical ones
\cite{MorPenTod98,BerGilJon98,KumSon00,Tod01,Son02,Ges04,SalCar06,GreWun06,Adl07}.  
The frictional alternative attracted less attention although the
special case (\ref{frSNEquad}) with the quadratic imaginary potential had been studied \cite{Has79} and was already 
utilized to generate pointer states longtime ago \cite{Dio87b}. That time, however, it was not quite clear to me 
why just frictional equations were better pointer state generators than others, and only after many years could
Kiefer and I analyze the complexity of the issue \cite{DioKie00}. Future investigation of the mathematical properties
of the frSNE is of interest. 

Finally, I would like to comment on the historic context. Penrose and myself got the same form (\ref{t_G}) of the 
decoherence-collapse time $t_G$. (Former discrepancies have been clarified in \cite{Dio05}.) I derived this equation
from the vNNE (\ref{vNNE}) and for quite a long time I did not see why Penrose did never buy the straightforward
construction of the vNNE from $t_G$. Later I understood his reservation. 
As mentioned in Sec.~\ref{sec_decoh}, too, the form (\ref{t_G}) yields a divergent $1/t_G$
for $R\rightarrow0$ and this divergence is inherited by the vNNE. This is indeed a serious problem and we do not exactly
know what to do with it, see \cite{Dio07,Dio05} and references therein. For point-like particles the decoherence speed $1/t_G$ 
would become infinite and the width of the wave function would tend to zero. But we do not expect the process of 
reduction to end up with singular wave functions rather we expect it ends up with localized ``pointer states'' of 
a certain finite width. Penrose defines the set of pointer states, independently of the equation
(\ref{t_G}) of $t_G$, as the ground state solitons of the SNE. So we have decent
pointer states even for point-like structures and we may get some hint how to cure the divergence of $1/t_G$.
I suggest warily: the hint might come from the stochastic wave equation (\ref{stWE}) which unifies the complete scenario 
of Fig.~2 in one equation. Section \ref{sec_match} invokes Ref.~\cite{Dio89} to claim the stochastic wave equation follows 
naturally from the Eq.~(\ref{t_G}) of $t_G$. The deterministic part of (\ref{stWE}) is regular for $R=0$, only its stochastic 
part is divergent. Shall gravity make a loophole somehow? We don't yet know. It is, nonetheless, remarkable that the 
deterministic part is the frictional Schr\"odinger---Newton equation. The frSNE has thus been in business already, and also 
independent arguments \cite{DioKie00} favor this class of {\it irreversible} modifications of the Schr\"odinger equation to 
generate robust pointer states, see also \cite{Gis84,Adl04}.   

\ack
This work was supported by the Hungarian OTKA Grant No. 49384.

\end{document}